\documentstyle[aps,prl,preprint,floats,epsfig]{revtex}

\begin{document}

\preprint{\tighten\vbox{\hbox{\hfil CLNS 01/1751}
                        \hbox{\hfil CLEO 01-16}
}}

\title{Branching Fraction and Photon Energy Spectrum for $b \rightarrow s \gamma$}  

% Your author list ***DOES NOT*** go here!
% is goes below where you are instructed to insert it...
\author{CLEO Collaboration}
% You will want to hard code the date once you are ready to submit your paper!
\date{\today}

\maketitle
\tighten

\begin{abstract} 
    We have measured the branching fraction and photon energy spectrum for the
radiative penguin process $b \rightarrow s \gamma$.  We find
${\cal B}(b \rightarrow s \gamma) = (3.21 \pm 0.43 \pm 0.27 ^{+0.18}_{-0.10})
\times 10^{-4}$, where the errors are statistical, systematic, and from theory
corrections.  We obtain first and
second moments of the photon energy spectrum above 2.0 GeV,
$\langle E_\gamma \rangle = 2.346 \pm 0.032  \pm 0.011\ \ {\rm GeV}$,
and
$\langle E^2_\gamma \rangle - \langle E_\gamma \rangle^2 = 0.0226  \pm 0.0066
\pm 0.0020  \ \ {\rm GeV}^2,$ where the errors are statistical and systematic.
From the first moment we obtain (in $\overline{MS}$, to order $1/M_B^3$ and
$\beta_0 \alpha_s^2$) the HQET parameter
$\bar \Lambda = 0.35  \pm 0.08  \pm 0.10 \ \ {\rm GeV}$.
\end{abstract}
\newpage

{
\renewcommand{\thefootnote}{\fnsymbol{footnote}}

% Insert author and address list here
\begin{center}
S.~Chen,$^{1}$ J.~W.~Hinson,$^{1}$ J.~Lee,$^{1}$
D.~H.~Miller,$^{1}$ V.~Pavlunin,$^{1}$ E.~I.~Shibata,$^{1}$ 
I.~P.~J.~Shipsey,$^{1}$
D.~Cronin-Hennessy,$^{2}$ A.L.~Lyon,$^{2}$ E.~H.~Thorndike,$^{2}$
% ...
\end{center}
 
\small
\begin{center}
$^{1}${Purdue University, West Lafayette, Indiana 47907}\\
$^{2}${University of Rochester, Rochester, New York 14627}\\
% ...
\end{center}

\setcounter{footnote}{0}
}
\newpage

% Insert body of the text here.
    As a result of several difficult
calculations\cite{SM-theory,Misiak-1,anatomy} the
Standard Model (SM)  expression for the $b \rightarrow s \gamma$ branching
fraction
has been determined to next-to-leading order.  The numerical value originally
given\cite{Misiak-1} was $(3.28 \pm 0.33)\times 10^{-4}$.  Recently,
Gambino and Misiak\cite{Misiak-2}  argue for use of a different choice for the
charm quark mass, and obtain $(3.73 \pm 0.30)\times 10^{-4}$.
The theoretical literature on
the connections between $b \rightarrow s \gamma$ and beyond-SM physics
is extensive\cite{bsm-theory}, including connections to SUSY, Technicolor,
charged Higgs, extra
dimensions, anomalous $W W \gamma$ couplings, dark matter, and more.
Thus a measurement different from the SM value would indicate beyond-SM physics,
and a measurement close to the SM value would constrain the parameters of SUSY,
Technicolor, and other beyond-SM physics options.

    In contrast, the photon energy spectrum is insensitive to beyond-SM Physics.
The photon mean energy $\langle E_\gamma \rangle$ is, to a good approximation,
equal
to half the $b$ quark mass $m_b$, while the mean square width of the energy
distribution, $\langle E^2_\gamma \rangle - \langle E_\gamma \rangle^2$, depends
on the mean square momentum of the $b$ quark within the $B$ meson.  A
knowledge of the $b$ quark mass and momentum allows a determination of the CKM
matrix element $V_{cb}$ from the $B$ semileptonic width, as shown in the
following Letter\cite{Moments}.  Further, the spectrum
provides information allowing a better determination of the CKM matrix element
$V_{ub}$ from the yield of leptons near the endpoint of $b \rightarrow u \ell
\nu$\cite{endpoint-theory,anatomy,btou}.

    CLEO's previously published measurement\cite{CLEO-old} of the
$b \rightarrow s \gamma$ branching fraction,
$(2.32 \pm 0.57 \pm 0.35)\times 10^{-4}$, was based on 3.0 ${\rm fb}^{-1}$ of
luminosity, on resonance plus off resonance.  The photon
energy spectrum then obtained was not sufficiently precise to be of interest.
Here we present a new study of $b \rightarrow s \gamma$, based on 9.1
${\rm fb}^{-1}$ on the $\Upsilon$(4S) resonance and 4.4 ${\rm fb}^{-1}$ at a
center-of-mass energy 60 MeV below the resonance (and below $B \bar B$
production threshold).  We present the branching
fraction and the first and second moments of the photon energy spectrum.  From
the first moment, we obtain\cite{Ligeti,Falkpc} the HQET parameter
$\bar \Lambda$ (physically, the energy of the light quark and gluon degrees of
freedom).

    The characteristic feature of the $b \rightarrow s \gamma$ decay is the high
energy photon, roughly monoenergetic, with $E_\gamma \approx m_b/2 \approx 2.3$
GeV.  In our previous measurement we considered the spectrum above 2.2 GeV,
incurring a model-dependent correction for the fraction of the spectrum below
2.2 GeV.  Here we use the spectrum down to 2.0 GeV, which includes 
$\sim$90\% of the $b \rightarrow s \gamma$ yield.

    The data used for this analysis were taken with the CLEO
detector\cite{detector} at the Cornell Electron Storage Ring (CESR), a symmetric
$e^+ e^-$ collider.  The CLEO detector measures charged particles over 95\% of
$4\pi$ steradians with a system of cylindrical drift chambers.  (For 2/3 of the
data used here, the innermost tracking chamber was a 3-layer silicon vertex
detector\cite{silicon}.)  Its barrel and endcap CsI electromagnetic calorimeters
cover 98\% of $4\pi$.  The energy resolution near 2.5 GeV in the central
angular region, $\vert \cos \theta_\gamma \vert < 0.7$, is 2.6\% r.m.s.,
including a low-side tail. Charged particle species ($\pi^\pm$, $K^\pm$,
$p /\bar p$) are
identified by specific ionization measurements ($dE/dX$) in the outermost drift
chamber, and by time-of-flight counters (ToF) placed just beyond the tracking
volume.  Muons are identified by their ability to penetrate the iron return yoke
of the magnet.  Electrons are identified by shower energy to momentum ratio
($E/P$), track-cluster matching, $dE/dX$, and shower shape.

We select hadronic events containing a high energy photon in
the central region of the calorimeter ($\vert \cos \theta_\gamma \vert <
0.7$).  The photon must not form a $\pi^0$ or $\eta$ with any other
photon in the event.

    There is a large background of high energy photons from continuum
processes: initial state radiation; photons from decays of $\pi^0$, $\eta$,
and other hadrons.  We use several techniques to suppress the continuum
background, measure what survives with the below-resonance data sample, and
subtract it from the on-resonance sample.

    There is also a significant background of photons from $B$ decay processes
other than $b \rightarrow s \gamma$, particularly in the lower portion of our
window, 2.0 -- 2.2 GeV.  We determine the dominant background sources, photons
from unvetoed $\pi^0 \rightarrow \gamma \gamma$ and
$\eta \rightarrow \gamma \gamma$, by directly measuring the $\pi^0$ and $\eta$
spectra, estimate the $K^0_L$ and $\bar n$ backgrounds using shower shape,
and estimate the remaining background sources via Monte Carlo 
techniques\cite{jana}.

Part of our continuum suppression comes from eight event shape
variables: normalized Fox-Wolfram\cite{Fox-Wolfram} second moment $R_2$,
$S_\perp$ (a measure of the momentum transverse to the photon
direction\cite{Jesse_thesis}), $R^\prime_2$ (the value of $R_2$ in the primed
frame,
the rest frame of $e^+ e^-$ following an assumed initial state radiation of the
high energy photon, with $R_2$ evaluated excluding the photon),
$\cos \theta^\prime$ ($\theta^\prime$ the angle, in the primed frame, between
the photon and the thrust axis of the rest of the event), and the energies in
$20^\circ$ and $30^\circ$ cones,
parallel and antiparallel to the high energy photon direction.  While no
individual variable has strong discrimination power, each posesses some.
Consequently, we combine the eight variables into a single variable $r$ which
tends towards +1 for $b \rightarrow s \gamma$ events and tends towards --1 for
continuum background events.  A neural network is used for this
task\cite{Jesse_thesis}.

    We obtain further continuum suppression from ``pseudoreconstruction'', and
from the presence of leptons.
In ``pseudoreconstruction'', we search events for combinations of particles that
reconstruct to a $B \rightarrow X_s \gamma$ decay.  For $X_s$ we use a
$K^0_S \rightarrow \pi^+ \pi^-$ or a charged track consistent with a $K^\pm$,
and 1 -- 4 pions, of which at most one may be
a $\pi^0$.  We calculate the candidate $B$ momentum $P$, energy $E$, and
beam-constrained mass $M \equiv \sqrt{E^2_{beam} - P^2}$.  A reconstruction
is deemed acceptable if it has $\chi^2_B < 20$, where
$$\chi^2_B \equiv \left( \frac{E - E_{beam}}{\sigma_E}\right)^2 + \left( \frac
{M - M_B}{\sigma_M}\right)^2\ \ ,$$
\noindent where, $\sigma_E = 40$ MeV, $\sigma_M = 4.0$ MeV.
If an event contains more than one acceptable reconstruction, the one with the
lowest $\chi^2_B$ is chosen.  (It is not important that the reconstruction be
correct in detail, as we are only using it for continuum suppression. It is correct 40\% of the time.)  

    For events with an acceptable reconstruction, we
add $\chi^2_B$, and $\vert \cos \theta_{tt} \vert$, where $\theta_{tt}$ is the
angle between the thrust axis of the candidate $B$ and the thrust axis of the
rest of the event, to the list of variables for continuum suppression.
If the event contains a lepton (muon or electron), then we add the momentum of
the lepton, $P_\ell$, and the angle between lepton and high energy photon,
$\theta_{\ell \gamma}$, to the list.
We thus have events with both a pseudoreconstruction and a lepton, for which we
use $r$, $\chi^2_B$, $\vert \cos \theta_{tt} \vert$, $P_\ell$, and
$\theta_{\ell \gamma}$ for continuum suppression; events with only a
pseudoreconstruction, for which we use $r$, $\chi^2_B$, and
$\vert \cos \theta_{tt} \vert$; events with only a lepton, for which we use
$r$, $P_\ell$, and $\theta_{\ell \gamma}$; and events with neither, for which we
use $r$.  In each of the four cases, the available
variables are combined using a neural net.  All nets are trained on signal and
continuum Monte Carlo.
We convert the output of each of the four nets,  $r_j$, into weights,
$w(r_j) = s(r_j)/[s(r_j) + (1 + a)b(r_j)]\ ,$ where  $s(r_j)$ is the expected
signal yield and $b(r_j)$ is the expected continuum background yield for that
value of the net output $r_j$, and $a$ is the luminosity scale
factor between on-resonance and off-resonance data samples ($\approx 2.0$).
Weights so defined minimize the statistical error on the $b \rightarrow s
\gamma$ yields after off-resonance subtraction, as described below.

    Were we to use the weights as just defined, the efficiency would be higher
for low mass $X_s$ states than for high mass $X_s$ states.  We thus modify the
weights for those cases where there is a
reconstruction, and thus a measurement of $M_{X_s}$, to render the efficiency
independent of the $X_s$ mass.  With these weights, the statistical error is
slightly worse, but a systematic error has been reduced.  The dependence of
efficiency on $M_{X_s}$ for events for which there is no pseudoreconstruction
remains.

    We sum weights on and off resonance, and subtract the off-resonance sum,
scaled by ${\cal L}/E^2_{cm}$ and with momenta scaled by the On/Off energy
ratio, from the on-resonance sum (the On-Off
subtraction).  Biases in this subtraction procedure have been estimated 
with Monte Carlo to be
(0.5 $\pm$ 0.5)\% of the subtraction; this has been applied as a correction and
as a contribution to the systematic error.
The on-resonance yields and scaled off-resonance yields are
shown, as a function of photon energy, in Fig.~1a, and tabulated in Table 1.
The On-Off subtracted yields are shown in Fig.~1b; they consist of a component
from $b \rightarrow s \gamma$, and a component from other $B$ decay processes.

We investigated the component from other $B$ decay processes with a $B \bar B$
Monte Carlo sample that included contributions from $b \rightarrow u$ and
$b \rightarrow s g$ processes, as well as the dominant $b \rightarrow c$ decay.
We found that the overwhelming source of background ($\sim$90\%) is photons from
$\pi^0$ or $\eta$ decay.  Consequently we carefully tuned the Monte Carlo to
match the data in $\pi^0$ and $\eta$ yields, and also $\eta^\prime$
and $\omega$.  We included radiative $\psi$ decays, 
$\rho \rightarrow \pi \gamma$, $a_1 \rightarrow \pi \gamma$, and final state
radiation.  We determine the background from neutral hadrons, $K^0_L$ and
$\bar n$, using information from the measured lateral distribution of the
showers.
The yield from $B$
decay processes is shown in Fig.~1b and tabulated in Table~1.
The error on this yield is dominated by the error on measured On-Off-
subtracted $\pi^0$ and $\eta$ yields, and by uncertainty in $K_L^0$ and 
$\overline{n}$ backgrounds from fits to shower shape.

    The fully subtracted spectrum, On - Off - other $B$ decay processes, is
shown in Fig.~2.  There is a clear peak near 2.3 GeV, indicative of
$b \rightarrow s \gamma$.  The region of interest for $b \rightarrow s \gamma$
is 2.0 -- 2.8 GeV.  The region above 3.0 GeV thus serves as a control,
indicating how well we have subtracted the continuum background, while
the region 1.5 -- 2.0 GeV serves as another control, indicating how well we have
accounted for the other $B$ decay processes.  The yields in both regions are
consistent with zero.  Subtracting 5\% more or less $B \bar B$ background
causes the $B$ background control region to have a deficit or excess, at the
1$\sigma$ level.  We take $\pm$5\% of the $B \bar B$ subtraction as the
systematic error on the subtraction.

    We determine the efficiency, weights per $b \rightarrow s \gamma$ event
generated with photon energy above 2.0 GeV, via Monte Carlo simulation.
To model the decay $b \rightarrow s \gamma$, {\it i.e.},
$B \rightarrow X_s \gamma$,
we use the spectator model of Ali and Greub\cite{Ali-Greub}, which includes
gluon bremsstrahlung and higher-order radiative effects.  Rather than
considering the Fermi momentum parameter $P_F$ and the $b$ quark average mass
$\langle m_b \rangle$ as quantities given from first principle, we treat them as
free parameters which allow us to vary the mean and width of the photon energy
spectrum.  We fit our measured spectrum over the range 2.0 -- 2.8 GeV, and use
the values of the parameters so obtained, and their errors, to define our model
of $B \rightarrow X_s \gamma$.  For hadronization of $s \bar q$ into $X_s$, we
have compared two approaches.  In the first, we have
taken several $K^*$ resonances and combined them in proportions to approximate
the $X_s$ mass distribution given by the Ali-Greub model.  In the second, we
have let JETSET hadronize the $s \bar q$, again giving the Ali-Greub $X_s$ mass
distribution.  We include in the systematic errors on the efficiency an
uncertainty in the modelling of the $B \rightarrow X_s \gamma$ decay ($P_F$,
$\langle m_b \rangle$, hadronization), an uncertainty in the simulation of
the detector performance (track-finding, photon-finding, resolutions), and an
uncertainty in the modelling of the other $B$.

    As an alternative to the Ali-Greub model for $B \rightarrow X_s \gamma$ decay,
we have used the description of Kagan and Neubert\cite{anatomy}, with their
simple
two-parameter shape function, with the two parameters $m_b$ and $a$ 
adjusted to 
fit our measured photon spectrum.  We obtain results very similar to those
obtained using Ali-Greub, both for efficiency and for the moments. (Because
we determine our  $B \rightarrow X_s \gamma$ model from a fit to data, our 
results are independent of the nominal description, as long as that 
description allows us two adjustable parameters to match data on mean and 
width.)

    To obtain the $b \rightarrow s \gamma$  branching fraction, we take the
yield between 2.0 and 2.7 GeV, 233.6 $\pm$ 31.2 $\pm$ 13.4 weights, where
the first error is statistical and the second is systematic, from 
$B \bar B$ ($\pm$ 5\%) and
continuum ($\pm$ 0.5\%) subtraction.  The efficiency is
$(3.93 \pm 0.15 \pm 0.17) \times 10^{-2}$ weights per event, where the first
error is from model dependence of the $B \rightarrow X_s \gamma$ decay and the
second error is from detector simulation and model dependence of the decay of
the other $B$.  Our sample contains 9.70 million $B \bar B$ pairs ($\pm$2\%).
We obtain an uncorrected branching fraction of
$(3.06 \pm 0.41 \pm 0.26) \times 10^{-4}$.

    This uncorrected branching fraction, what we directly measure, is the
branching fraction for $b \rightarrow s \gamma$ plus $ b \rightarrow d \gamma$,
for $B$-rest-frame photon energies above 2.0 GeV.  We apply two ``theory''
corrections to obtain a branching fraction of more direct interest.
    Using a model for $b \rightarrow d \gamma$ similar to that used for
$b \rightarrow s \gamma$, we find the efficiency for $b \rightarrow d \gamma$ to
be the same as that for $b \rightarrow s \gamma$. (Pseudoreconstruction favors $b \rightarrow s \gamma$ over $b \rightarrow d \gamma$ but shape variables and
photon energy spectrum favor $b \rightarrow d \gamma$ over 
$b \rightarrow s \gamma$.)  The SM expectation is that
the $b \rightarrow d \gamma$ and $b \rightarrow s \gamma$ branching fractions
are in the ratio $\vert V_{td}/V_{ts} \vert ^2$.  Using
$\vert V_{td}/V_{ts} \vert = 0.20 \pm 0.04$\cite{PDG} (taking 
$V_{ts}$ = - $V_{cb}$, from unitarity), we correct the branching
fraction down by (4.0 $\pm$ 1.6)\% of itself, to remove the
$b \rightarrow d \gamma$ contribution.
    The fraction of $b \rightarrow s \gamma$ decays with photon energies above
2.0 GeV is sensitive to the $b$ mass and Fermi momentum.  We use
$0.915 ^{+0.027}_{-0.055}$, as given by Kagan and Neubert\cite{anatomy,extrap} to
extrapolate our
branching fraction to the full energy range (actually, to energies above
0.25 GeV).
    With these two corrections, we have
$${\cal B}(b \rightarrow s \gamma) = (3.21 \pm 0.43 \pm 0.27 ^{+0.18}_{-0.10})
\times 10^{-4}\ ,$$
\noindent for the branching fraction for $b \rightarrow s \gamma$ alone, over
all energies.  The first error is statistical, the second systematic, the third
from the theory corrections.
This result is in good agreement with the Standard Model prediction.  It is also
in agreement with our previously-measured result\cite{CLEO-old}, which it
supercedes.

    We have calculated the first and second moments of the photon energy
spectrum, in two ways.  In the first, we correct the raw spectrum for the energy
dependence of the efficiency, calculate moments from that spectrum, apply
corrections for experimental resolution and for the transformation from lab
frame to $B$ rest frame.  We then apply a final, empirical correction, obtained
by carrying out the above procedure on Monte Carlo.
In the second way, we take our best fit Monte Carlo model, with $P_F$
and $\langle m_b \rangle$ determined from a fit to the spectrum, and take the
moments given by that model.  The two ways of obtaining moments agree, and we take their average.  We thus
obtain moments in the $B$ rest frame, for $E_\gamma(rest frame) > 2.0$ GeV:

$$\langle E_\gamma \rangle = 2.346 \pm 0.032  \pm 0.011 \ \ {\rm GeV}\ .$$
$$\langle E^2_\gamma \rangle - \langle E_\gamma \rangle^2 = 0.0226  \pm 0.0066
\pm 0.0020 \ \ {\rm GeV}^2\ .$$

    Heavy Quark Effective Theory and the Operator Product Expansion, with the
assumption of parton-hadron duality, allows
inclusive observables to be written as double expansions in powers of
$\alpha_s$ and $1/M_B$. The expressions \cite{Ligeti},\cite{Falkpc} for the
moments of the photon energy spectrum in $B \rightarrow X_s \gamma$, for
$E_\gamma > 2.0$ GeV, in the $\overline{MS}$ renormalization scheme,
to order $\beta_0 \alpha^2_s$ and $1/M_B^3$, are given
in Eqs.~\ref{eq:egammom1} and \ref{eq:egammom2}.

\begin{eqnarray}
\langle E_{\gamma}\rangle = & \frac{M_B}{2}
[1-.385 \frac{\alpha_s}{\pi}-.620 \beta_0 (\frac{\alpha_s}{\pi})^2 -
\frac{\bar \Lambda}{M_B} (1-.954 \frac{\alpha_s}{\pi} -
1.175 \beta_0 (\frac{\alpha_s}{\pi})^2) \nonumber \\
& - \frac{13 \rho_1 - 33 \rho_2}{12 M_B^3}  - \frac{{\cal T}_1 + 3
{\cal T}_2 + {\cal T}_3 +3 {\cal T}_4}{4 M_B^3}
- \frac{\rho_2 C_2}{9 M_B M_D^2 C_ 7}
 + {\cal O}(1/M^4_B)],
\label{eq:egammom1}
\end{eqnarray}

\begin{eqnarray}
&\langle (E_{\gamma} - \langle E_{\gamma} \rangle)^2 \rangle =
M_B^2 [{-\lambda_1 \over {12 }{M_B^2}} + (0.00815\frac{\alpha_s}{\pi}+
0.01024 \beta_0 (\frac{\alpha_s}{\pi})^2) \nonumber \\
&- \bar \frac{\Lambda}{M_B} (0.05083\frac{\alpha_s}
{\pi}+0.05412\beta_0(\frac{\alpha_s}{\pi})^2)
- \frac{2 \rho_1 - 3 \rho_2}{12 M_B^3}  
- \frac{ {\cal T}_1 + 3 {\cal T}_2}{12 M_B^3}
 + {\cal O}(1/M^4_B)],
\label{eq:egammom2}
\end{eqnarray}

\noindent where $C_2$ and $C_7$ are Wilson coefficients and $\beta_0$
is the one-loop QCD $\beta$ function.
The $1/M_B^3$ parameters $\rho_i$,
${\cal T}_i$ are estimated, from dimensional considerations,  to be
$\sim(0.5 {\rm GeV})^3$.  Using Eq.~\ref{eq:egammom1}, we obtain

$$\bar \Lambda = 0.35 \pm 0.08 \pm 0.10\ {\rm GeV}\ ,$$

\noindent where the first error is from the experimental error in the
determination of the first moment, and the second error is from the theoretical
expression, in particular from the neglected $1/M_B^3$ terms, and the uncertainty
of the scale to use for $\alpha_s$, which we take to be from $m_b/2$ to
$2 m_b$.

    The expression for the second moment converges slowly in $1/M_B$, and so we
make no attempt to extract expansion parameters from it.

    In summary, we have measured the branching fraction for $b \rightarrow s
\gamma$ to be $(3.21 \pm 0.53) \times 10^{-4}$, in good agreement with the
Standard Model expectation.  We have measured the first and second moments of
the $B$ rest frame photon energy spectrum above 2.0 GeV to be
$\langle E_\gamma \rangle = 2.346 \pm 0.034  \ \ {\rm GeV}\ ,$ and
$\langle E^2_\gamma \rangle - \langle E_\gamma \rangle^2 = 0.0226  \pm 0.0069
\ \ {\rm GeV}^2\ .$  From the first moment, we have obtained a value for the
HQET OPE parameter 
$\bar \Lambda = 0.35 \pm 0.08 \pm 0.10\ {\rm GeV}\ .$

We gratefully acknowledge the effort of the CESR staff in providing us with
excellent luminosity and running conditions.
We thank A. Falk, A. Kagan and M. Neubert for many informative conversations
and correspondences.
This work was supported by 
the National Science Foundation,
the U.S. Department of Energy,
the Research Corporation,
and the Texas Advanced Research Program.

\begin{table}
\begin{center}
\begin{tabular}{|l|c|c|c|}
Energy Range  &1.5 -- 2.0 GeV & 2.0 -- 2.7 GeV & 3.0 -- 5.0 GeV  \\ \hline
On       & 2718.8$\pm$ 17.2 & 1861.7$\pm$ 16.5 & 1083.2$\pm$ 8.0 \\
$a\times$Off& 1665.6$\pm$ 14.6& 1399.4$\pm$  15.31& 1101.4$\pm$ 11.6\\
On$-a\times$Off& 1053.2$\pm$ 22.6& 462.3$\pm$ 22.5&
--18.2$\pm$ 14.1\\ \hline
$B$ backgrounds & 1033.0$\pm$ 35.4 & 228.6$\pm$ 21.6 &
1.5 $\pm$ 1.1 \\ \hline
On $-a\times$Off $-B$ & 20.2$\pm$42.0 & 233.6$\pm$31.2
& --19.8$\pm$ 14.2 \\
\end{tabular}
\end{center}
\caption{ Yields (summed weights) with statistical errors for three photon
energy intervals.  Given are yields On $\Upsilon(4S)$ resonance,
scaled Off-resonance
yields, On minus scaled Off, estimated backgrounds from $B$ decay processes
other than $b \rightarrow s \gamma$ and $b \rightarrow d \gamma$, and On minus
scaled Off minus $B$ backgrounds, the putative $b \rightarrow s \gamma$ plus
$b \rightarrow d \gamma$ signal.}
\end{table}

\begin{figure}
\begin{center}
\epsfxsize=3.25in
\epsfbox{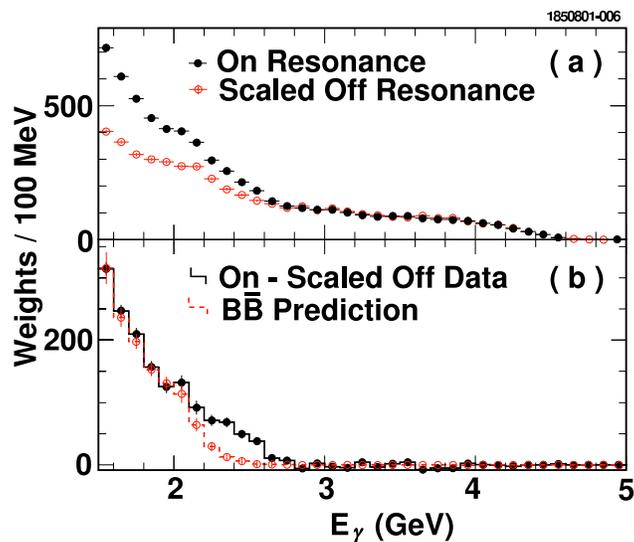}
\hfill \caption{
Photon energy spectra (weights per 100 MeV).  The upper plot (a) shows the On
$\Upsilon(4S)$ and scaled Off-resonance spectra.  The lower plot (b) shows their
difference, and the spectrum estimated for $B$ decay processes other than
$b \rightarrow s \gamma$ and $b \rightarrow d \gamma$.}
\end{center}
\end{figure}

\begin{figure}
\begin{center}
\epsfxsize=3.25in
\epsfbox{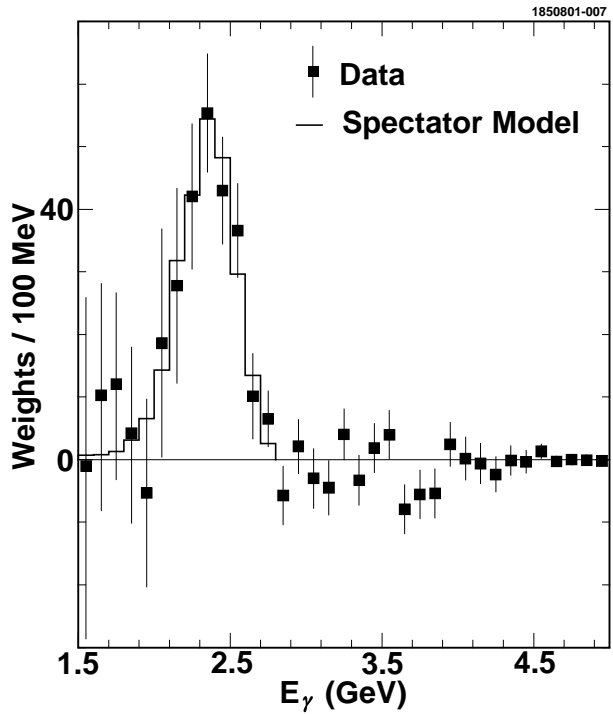}
\hfill \caption{
Observed laboratory frame photon energy spectrum (weights per 100 MeV) for
On minus scaled Off minus $B$ backgrounds, the putative
$b \rightarrow s \gamma$ plus $b \rightarrow d \gamma$ signal.  No corrections
have been applied for resolution or efficiency.  Also shown is the
spectrum from Monte Carlo simulation of the Ali-Greub spectator model with
parameters $\langle m_b \rangle$ = 4.690 GeV, $P_F$ = 410 MeV/c, a good fit to
the data.
}
\end{center}
\end{figure}


\begin{thebibliography}{99}

\bibitem{SM-theory} K. Adel and Y. P. Yao, Phys. Rev. D {\bf 49}, 4945 (1994);
C. Greub and T. Hurth, Phys. Rev. D {\bf 56}, 2934 (1997);
A. Ali and C. Greub, Phys. Lett. {\bf B 361}, 146 (1995);
C. Greub, T. Hurth, and D. Wyler, Phys. Lett. {\bf B 380}, 385 (1996); Phys.
Rev. D {\bf 54}, 3350 (1996);
A. Czarnecki and W. J. Marciano, Phys. Rev. Lett. {\bf 81}, 277 (1998).

\bibitem{Misiak-1} K. Chetyrkin, M. Misiak, and M. M\"{u}nz, Phys. Lett.
{\bf B 400}, 206 (1997).

\bibitem{anatomy} A. L. Kagan and M. Neubert, Eur. Phys. J. {\bf C 7}, 5 (1999).

\bibitem{Misiak-2} P. Gambino and M. Misiak, hep-ph/0104034.

\bibitem{bsm-theory} SPIRES lists over 500 citations to our earlier
$b \rightarrow s \gamma$ branching fraction paper (ref.\cite{CLEO-old}), most
of them theory papers addressing beyond-SM issues.

\bibitem{Moments} D. Cronin-Hennessy {\it et al.} (CLEO), CLEO 01-17, CLNS
01/1752, submitted to Phys. Rev. Lett.

\bibitem{endpoint-theory} M. Neubert, Phys. Rev D {\bf 49}, 4623 (1994);
A. K. Leibovich, I. Low, and I. Z. Rothstein,
Phys. Rev. D {\bf 61}, 053006 (2000).

\bibitem{btou} F. De Fazio and M. Neubert, JHEP06, 017 (1999).

\bibitem{CLEO-old}  M. S. Alam {\it et al.} (CLEO),
Phys. Rev. Lett. {\bf 74}, 2885 (1995).

\bibitem{Ligeti} Z. Ligeti, M. Luke, A. V. Manohar, and M. B. Wise,
Phys. Rev. D {\bf 60} 034019, (1999); C. Bauer, 
Phys. Rev. D {\bf 57}, 5611. (1998).

\bibitem{Falkpc}  Adam Falk and Zoltan Ligeti, private communications.

\bibitem{detector} Y. Kubota {\it et al.} (CLEO), Nucl. Instrum. Meth. A {\bf
320}, 66 (1992).

\bibitem{silicon} T. Hill, Nucl. Instrum. Methods Phys. Res., Sect. A {\bf 418},
32 (1998).

\bibitem{jana} Jana Thayer, Ph.D. thesis, Ohio State University (in progress).

\bibitem{Fox-Wolfram}  G. Fox and S. Wolfram, Phys. Rev. Lett. {\bf 41}, 1581
(1978).

\bibitem{Jesse_thesis}  J. A. Ernst, Ph.D. thesis, Univ. of Rochester (1995).

\bibitem{Ali-Greub} A. Ali and C. Greub, Phys. Lett. {\bf B 259}, 182 (1991);
and private communications.

\bibitem{PDG}  D. E. Groom {\it et al.} (PDG), Eur. Phys. J.
C {\bf 15}, 1 (2000).

\bibitem{extrap} M. Neubert, hep-ph/9809377.

\end{thebibliography}
\end{document}